\documentclass[twocolumn, letterpaper]{article}
\usepackage[margin=1in]{geometry}

\usepackage[backend=biber]{biblatex}
\usepackage{graphicx}
\usepackage{subfigure}
\usepackage[colorlinks, hidelinks]{hyperref}

\usepackage{orcidlink}

\title{Improvements \& Evaluations on the \\ MLCommons CloudMask Benchmark}

\author{Varshitha Chennamsetti, Laiba Mehnaz, Dan Zhao, Banani Ghosh, 
Sergey V. Samsonau \orcidlink{0000-0002-0835-2970} \\
AI for Scientific Research \\ 
New York University, New York, NY 10003, USA}
\date{}

\begin{document}

\twocolumn[
\begin{@twocolumnfalse}
  \maketitle
  \begin{abstract}
    In this paper, we report the performance benchmarking results of deep learning models on MLCommons' Science cloud-masking benchmark using a high-performance computing cluster at New York University (NYU): NYU Greene.  MLCommons is a consortium that develops and maintains several scientific benchmarks that can benefit from developments in AI. We provide a description of the cloud-masking benchmark task, updated code, and the best model for this benchmark when using our selected hyperparameter settings. Our benchmarking results include the highest accuracy achieved on the NYU system as well as the average time taken for both training and inference on the benchmark across several runs/seeds.   Our code can be found on GitHub. MLCommons team has been kept informed about our progress and may use the developed code for their future work.
    \vspace{0.5em}
    
    \vspace{0.5em}
    \vspace{0.5em}
  \end{abstract}
  \end{@twocolumnfalse}%
]

\section{Introduction}

With artificial intelligence (AI) becoming a powerful technology transforming almost every aspect of life, several scientific niches/fields are yet to fully benefit from advances in AI. As such, there is a need for an initiative to increase awareness and push forward innovation in scientific fields where the potential for AI is huge but under-explored. MLCommons \cite{www-mlcommons-research}  is one such community effort to promote scientific AI benchmarking. The MLCommons Science Working Group \cite{Thiyagalingam2022AIBF} has so far developed four scientific benchmarks in four varying fields: atmospheric sciences, solid-state physics, healthcare, and earthquake forecasting. In this paper, we present our preliminary work on the cloud masking science benchmark in the atmospheric sciences domain. 

The objective of this benchmark is to accurately identify cloud pixels given a satellite image (image segmentation). The European Space Agency (ESA) \cite{www-sentinel92} has deployed a series of satellites to monitor the global environment. Among these satellites, Sentinel-3 focuses on ocean surface topography, sea surface temperature (SST), and land surface temperature (LST). The process of retrieving SST and LST normally begins with cloud screening/masking, followed by the actual temperature estimation of the pixels that do not contain clouds. Cloud masking is essential since the presence of clouds can distort temperature estimation, making it a crucial step in accurate SST and LST estimation.



Several methods have been used for this task ranging from simpler rule-based methods \cite{Saunders1986AnAS,Saunders1988AnIM,Merchant2005ProbabilisticPB, Zhu2012ObjectbasedCA} to deep learning techniques \cite{Li2019DeepLB,Domnich2021KappaMaskAC,Yan2018CloudAC,WIELAND2019111203,JEPPESEN2019247}. Some examples of the rule based techniques have included thresholding \cite{Saunders1986AnAS,Saunders1988AnIM} and Bayesian masking \cite{Merchant2005ProbabilisticPB}. Bayesian masks rely on Bayes theorem and prior meteorology information/data to provide probabilities for each pixel having a cloud or not. Using these probabilities and applying them to a given image, a cloud mask can be generated. On the other hand, deep learning methods \cite{Li2019DeepLB,Domnich2021KappaMaskAC,Yan2018CloudAC,WIELAND2019111203,JEPPESEN2019247} use computer vision models and treat the task image segmentation in order to learn and generate/predict these cloud masks. 


As part of the cloud masking benchmark, we are provided with the satellite images from Sentinel-3 for cloud masking. In this paper, we present our work on this benchmark using the reference/baseline implementation as well as a U-Net model \cite{Ronneberger2015UNetCN}. Our evaluation attempts to be holistic: our experiments benchmark both task or model performance in terms of accuracy on the cloud masking tasks as well as computational performance and scalability in terms of training and inference time. We conduct our analysis and experiments on New York University's (NYU) High Performance Computing (HPC) cluster, HPC Greene.

\begin{table*}[!ht]
    \centering
    \caption{This table presents the several methods used for cloud masking with their respective dataset, ground truth, and performance.}
    \label{tab:datasets}
    \resizebox{2\columnwidth}{!}{
    \begin{tabular}{|l|c|l|l|l|l|}
    \hline
        {\bf} & {\bf Reference} & {\bf Dataset} & {\bf Ground-truth} & {\bf Model}  & {\bf Accuracy} \\ \hline
        1 & \cite{Merchant2005ProbabilisticPB} & ATSR-2 & Human annotation & Bayesian screening & 0.917\\ \hline
        2 & \cite{WIELAND2019111203} & Sentinel-2 & Software-assisted human annotation (QGIS) & U-Net & 0.90 \\ \hline
        3 & \cite{WIELAND2019111203} & Landsat TM & Software-assisted human annotation (QGIS) & U-Net & 0.89 \\ \hline
        4 & \cite{WIELAND2019111203} & Landsat ETM+ & Software-assisted human annotation (QGIS) & U-Net & 0.89 \\ \hline
        5 & \cite{WIELAND2019111203} & Landsat OLI & Software-assisted human annotation (QGIS) & U-Net & 0.91 \\ \hline
        6 & \cite{Domnich2021KappaMaskAC} & Sentinel 2 & Software-assisted human annotation (CVAT) & KappaMask & 0.91 \\ \hline
        7 & \cite{JEPPESEN2019247} & Landsat 8 Biome and SPARCS & Human annotation & RS-Net & 0.956 \\ \hline
    \end{tabular}}

\end{table*}

\section{MLCommons}

Scientific benchmarking is different from standard AI/ML benchmarking in that the underlying datasets can differ in terms of its volume (e.g., the amount/scale of data) and type (e.g., sensor data, multi-channel images). As a result, this can require significant computational resources such as supercomputers  \cite{Farrell2021MLPerfHA} due to the I/O impact of large-scale datasets. MLCommons is a consortium that oversees several scientific benchmarks in various scientific domains that can benefit from developments in AI. One of the working groups in MLCommons is the Science Working Group \cite{Thiyagalingam2022AIBF}, which, at the time of writing, manages four different benchmarks. Each benchmark includes the relevant data for training and testing purposes along with a defined metric by which task performance is measured and assessed. In our case, the cloud masking benchmark contains 180 GB worth of satellite image data with the task of identifying cloud cover masks, whereas the relevant metrics include classification accuracy and scalability (e.g., over compute time and other resources like GPUs). 
For each benchmark, there is a reference/baseline implementation provided by the Science Working Group. We use reference implementation of this benchmark for this paper.

\section{Related Work}

Cloud masking is a crucial task that is well-motivated for meteorology and applications in environmental sciences. Given satellite images, cloud masks are generated such that each pixel is labeled to either have cloud or not. To produce these binary cloud masks, the traditional approaches have been: thresholding \cite{Saunders1986AnAS,Saunders1988AnIM} and Bayesian \cite{Merchant2005ProbabilisticPB} methods. Thresholding methods consist of several threshold tests where spectral and spatial properties of the images are compared with ranges that are believed to be indicative of a {\em clear-sky} pixel. And those other pixels that are not labeled as {\em clear-sky} are flagged as {\em cloudy} pixel. This method was widely used from the late 1980s to the early 2000s \cite{Merchant2005ProbabilisticPB}. The gradual transition away from threshold tests was due to its long-criticized limitations: firstly, threshold settings rely on domain expertise about indicators of cloudiness that may not be objective, which also makes later modification and updates difficult; secondly, thresholds provide users no flexibility in the trade-off between SST coverage and SST bias; third, threshold tests do not make use of all available prior information. These shortcomings of thresholding methods are improved by later developed Bayesian methods \cite{Merchant2005ProbabilisticPB}.

Bayesian methods deduce the probability of a clear sky over each pixel by applying Bayes' theorem. As a result, these Bayesian approaches are fully probabilistic and deduce pixels based on prior information and observations in images. Compared to threshold tests, Bayesian methods achieve better accuracy in predicting pixels' cloudiness, offering generality and conceptual clarity in its approach as well as enhancing maintainability and adaptability \cite{Merchant2005ProbabilisticPB}. 

More recently, the rise of deep learning has led to the use of convolutional neural networks (CNNs) for generating these cloud masks (Table \ref{tab:datasets}). CNNs have achieved superior performance due to their superior abilities for automatic feature extraction, among other aspects. \cite{WIELAND2019111203} have shown how a CNN for image segmentation, the U-Net, can be used to generate cloud masks to better performance than Fmask  \cite{Zhu2012ObjectbasedCA}, a state-of-the-art rule-based approach.\cite{Zhu2012ObjectbasedCA}. \cite{JEPPESEN2019247} introduced Remote Sensing Network (RS-Net), an architecture based on U-Net for cloud masking that has been shown to achieve higher performance compared to Fmask. Along similar lines, KappaMask \cite{Domnich2021KappaMaskAC} is another U-Net based CNN model that outperformed the rule-based Fmask algorithm. MSCFF \cite{Li2019DeepLB} is a CNN model that uses an encoder-decoder model to extract high-level features that also outperform Fmask on several satellite datasets. All these models have reported their performances on several satellite images such as Sentinel-2, Landsat, etc., and also make use of human-annotated (some assisted by software) ground truth values. On the contrary, the cloud-masking benchmark makes use of Sentinel-3's images and uses Bayesian masks as the ground truth. To our best knowledge, there is no previous work done using Sentinel-3 images except the reference implementation provided by MLCommons Science Working Group that achieves 92\% classification accuracy on the test set.

\begin{figure*}[ht]
\centering\includegraphics[width=0.75\paperwidth]{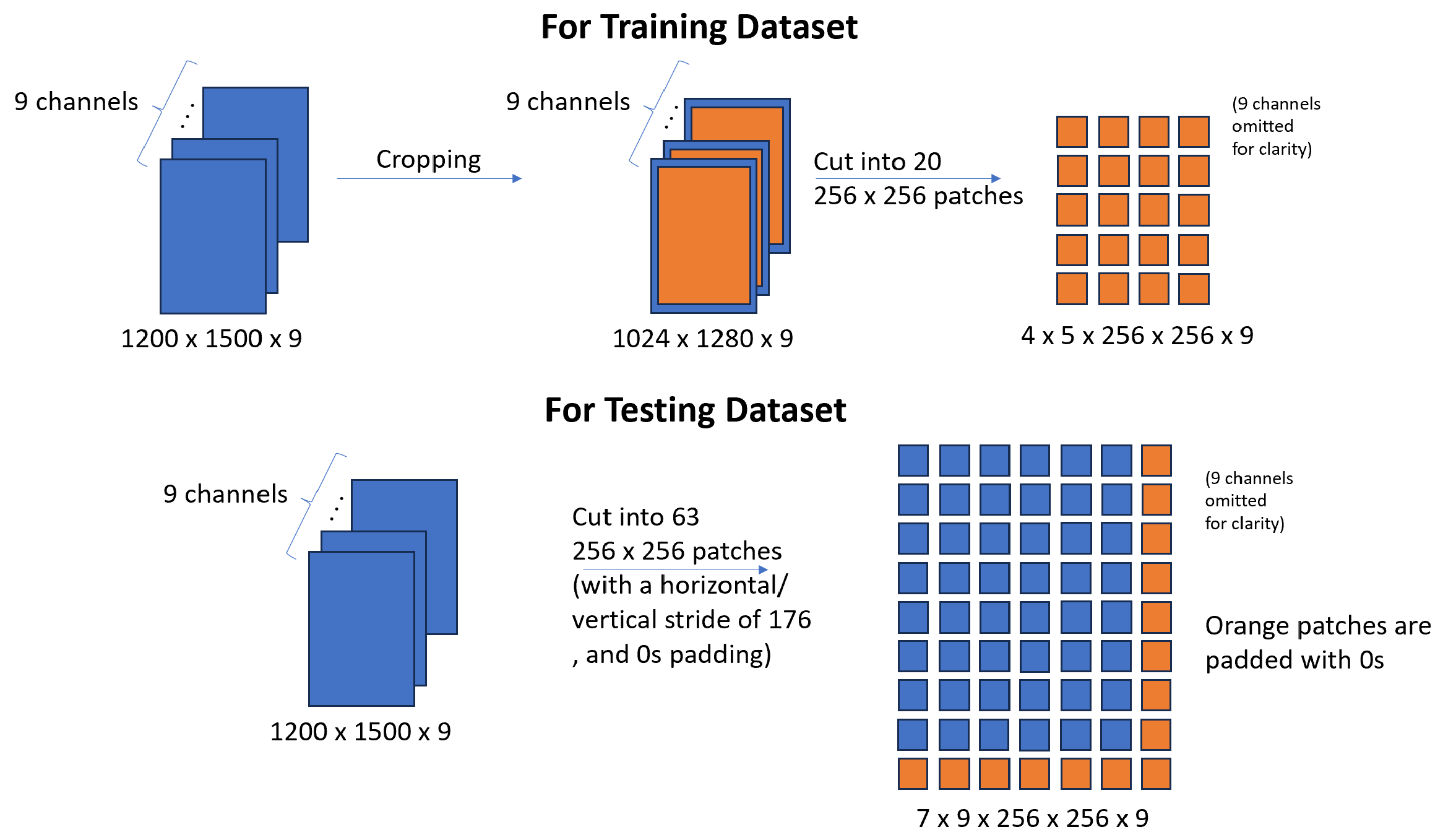}
\caption{An illustration of how training and testing datasets are pre-processed before training/inference.}
\label{fig:preprocessing}
\end{figure*}

\begin{table*}[!h]
    \caption{Overview of the computing resources used \cite{las23-cloudmask}, as of Spring 2023.}
    \label{tab:hwoverview}
    \begin{center}
    \begin{tabular}{|l|r|r|r|r|r|r|r|}
        \hline
            {\bf Machine}  & {\bf Cores} & {\bf Memory} & {\bf GPU}   &   {\bf Memory} & {\bf GPUs} & {\bf Nodes}  & {\bf Commissioned} \\ 
                     &  {\bf /Node} & {\bf /Node}  &  {\bf Type}  & {\bf /GPU (GB)}     &   {\bf /Node}        & & \\
        \hline
        \hline
         Greene (NYU)	& 48	& 384	& RTX8000 	& 48	& 4	& 73 & Nov 2020\\
	        & 48	& 384	& V100 	    & 32	& 4	& 10 & other dates \\
	        & 40	& 512	& V100 	    & 3     & 8 & 1  & not published \\
	        & 40	& 384	& V100 	    & 32    & 4	& 1  & \\
	        & 40	& 384	& V100 	    & 16    & 4	& 6  & \\
	        & 80	& 1024	& A100  8358	& 80    & 4 & 31 & \\
	        & 64	& 512	& A100  8380	& 80    & 4 & 9  & \\

         \hline
         \hline
    \end{tabular}
    \end{center}

\end{table*}

\section{Dataset} 

As mentioned in the previous Section, the cloud-masking benchmark provides 180GB worth of Sentinel-3 satellite images for this task. The dataset consists of a total of 1070 images, both captured at day and night. The dataset comes with the train-test split where $90\%$ is used for training, and $10\%$ is used for testing. 
The images are of the dimensions 1200 x 1500 with fifteen different channels. Three channels are used to represent brightness, six channels are used to represent reflectance, and six channels are used to represent radiance. However, for the benchmark, as provided in the reference implementation, only a total of nine channels, i.e., six channels of reflectance and three channels of brightness are used. 

For data pre and post-processing, per the benchmark's reference implementation, the input images and their ground truths are first cropped and then divided into smaller-sized $256 x 256$ patches, keeping all the nine channels intact. After creating these patches of individual images, we have a total of 19,400 images for training. Hereafter, we refer to patches as images. These images are then split into training and validation sets using a 80 to 20 split ratio.

For the test set, the images are neither cropped nor shuffled. Overlapping smaller images (patches) of size $256 \times 256$ are created both for the input image and not for the ground truth for a total of $6,300$ images. After obtaining the predictions from the model, these  $256 \times 256$ images are reconstructed to the original size and then evaluated with the ground truth. These data processing steps are shown in Figure \ref{fig:preprocessing}.

During training, the model takes an image of size $256 \times 256$, and generates a cloud mask of the same size. Once the cloud masks have been generated by the model during training, the accuracy is reported as the number of pixels that are correctly classified (as cloud or not). 

\begin{table*}[!h]
    \centering
    \caption{Training and test statistics (train/test accuracies and times) across five training runs of our U-Net on V100 GPUs at NYU Greene. The average training/inference times per epoch are calculated as the average amount of time per epoch and are reported in minutes. The number of training epochs is determined either by convergence of train/test accuracies or via early stopping on the best loss/accuracy within our patience hyper-parameter (25 epochs).}
    \begin{tabular}{|c|c|c|c|c|}
    \hline
        \bf{Epochs} &  \bf{Train Acc.} & \bf{Test Acc.} & \bf{Avg. Train Time } & \bf{Avg. Inference Time} \\ \hline
        
        200 &   0.896 & 0.884 & 142.47 & 1.73\\ \hline        
        200 &  0.897 & 0.890 & 126.44 & 1.53 \\ \hline        
        183 &   0.897 & 0.885 & 121.21 & 1.41 \\ \hline      
        162 &   0.897 & 0.890 & 108.16 & 1.44\\ \hline        
        147 &   0.909 & 0.896 & 98.75 & 1.46\\ \hline        
    \end{tabular}
    \label{tab: Results}
\end{table*}


\begin{figure*}[!h]
   \centering
   \begin{minipage}{.45\textwidth}
      \subfigure[Training Loss]{
         \includegraphics[width=0.9\textwidth]{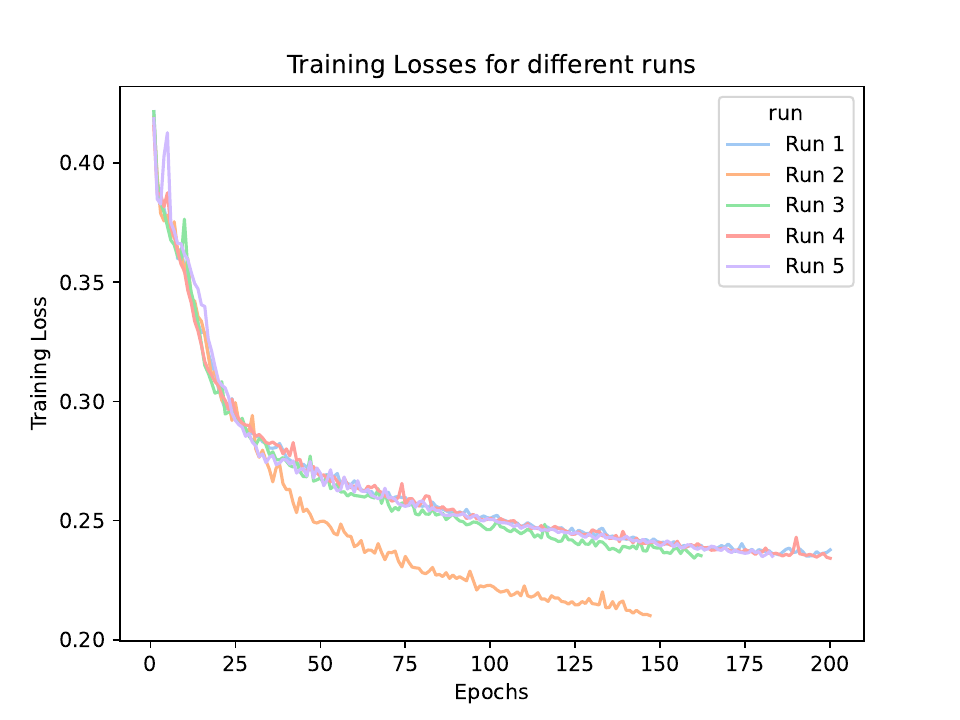}
      }
   \end{minipage}
   \hfill
   \begin{minipage}{.45\textwidth}
      \subfigure[Training Accuracy]{
         \includegraphics[width=0.9\textwidth]{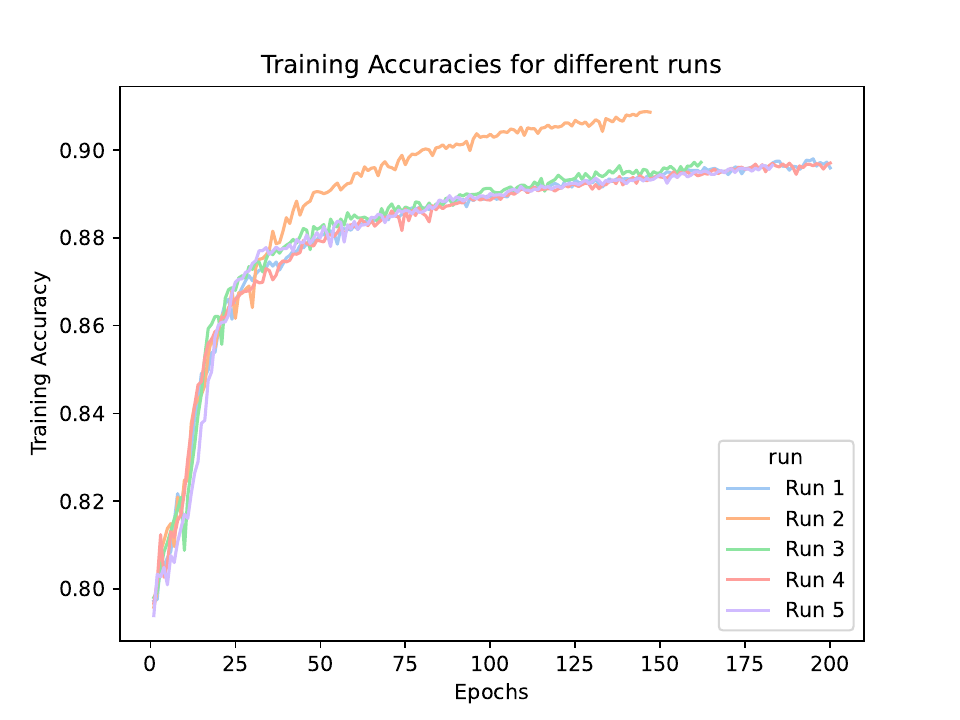}
      }
   \end{minipage}
   \caption{Training set results with different runs. With early stopping and patience of 25, the 5 different runs stop their training and save the model weights at epochs 200, 147, 162, 200, 183, respectively.}
   \label{fig:all_runs_tr}
\end{figure*}



\begin{figure*}[!h]
   \centering
   \begin{minipage}{.45\textwidth}
      \subfigure[Validation Loss]{
         \includegraphics[width=0.9\textwidth]{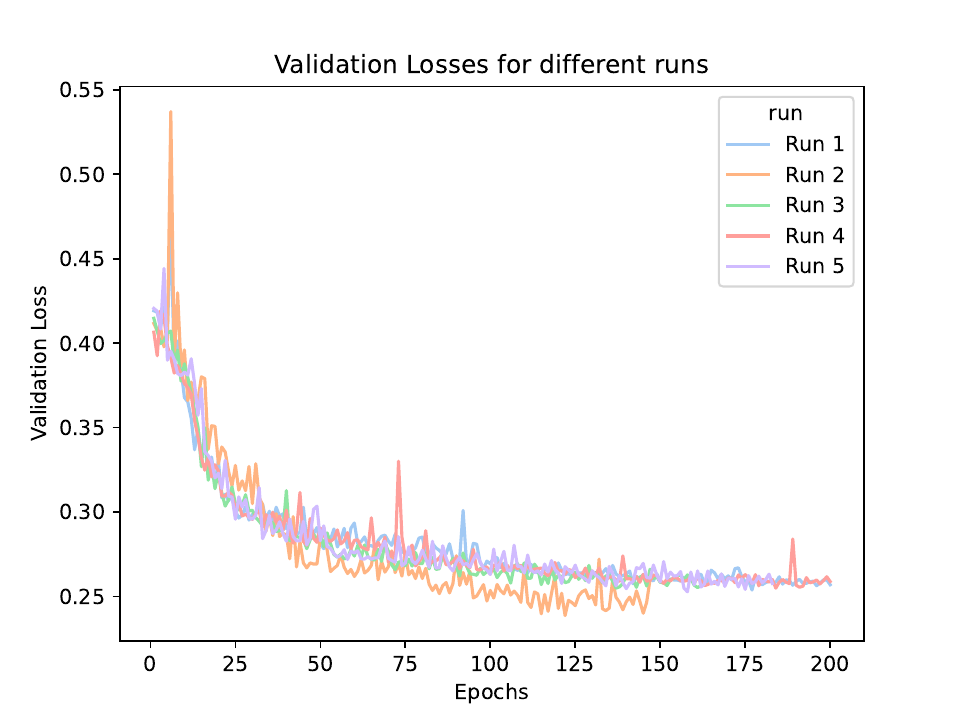}
      }
   \end{minipage}
   \hfill
   \begin{minipage}{.45\textwidth}
      \subfigure[Validation Accuracy]{
         \includegraphics[width=0.9\textwidth]{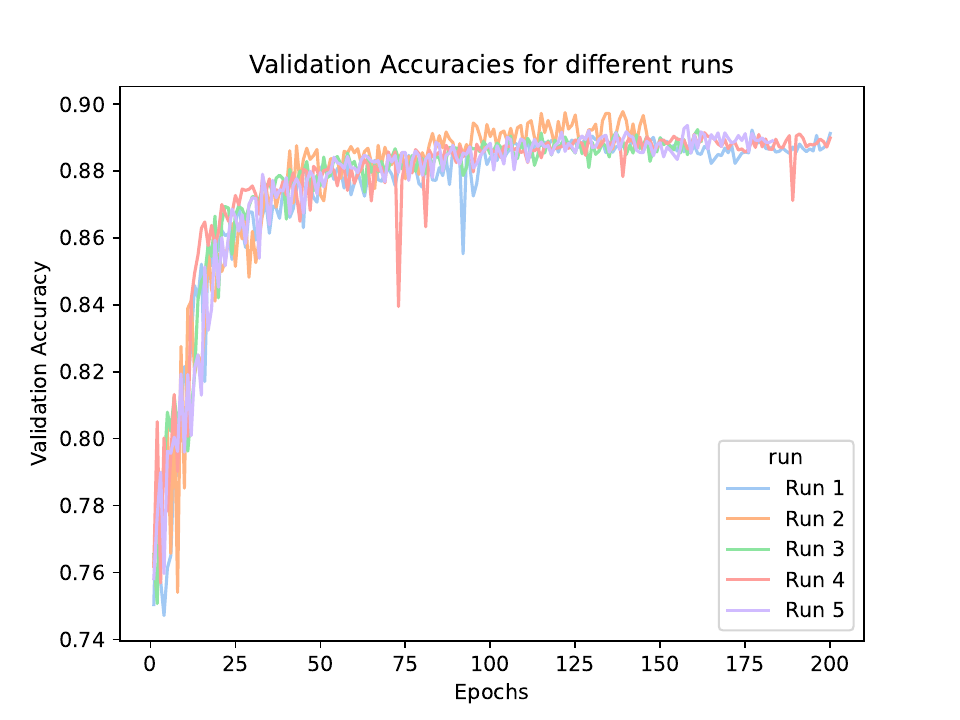}
      }
   \end{minipage}
   \caption{Validation set results across different runs. With early stopping and patience of 25, the 5 different runs stop their training and save the model weights at epochs 200, 147, 162, 200, 183, respectively.}
   \label{fig:all_runs_val}
\end{figure*}

\begin{figure*}[!h]
   \centering
   \begin{minipage}{.45\textwidth}
      \subfigure[Training and Validation Loss]{
         \includegraphics[width=0.9\textwidth]{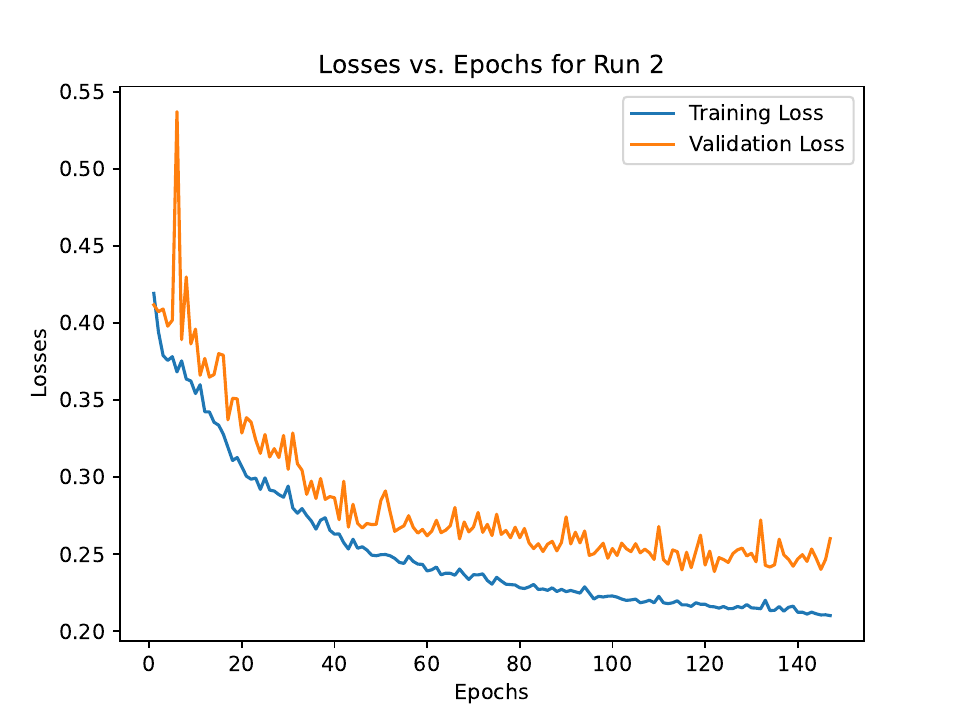}
      }
   \end{minipage}
   \hfill
   \begin{minipage}{.45\textwidth}
      \subfigure[Training and Validation Accuracy]{
         \includegraphics[width=0.9\textwidth]{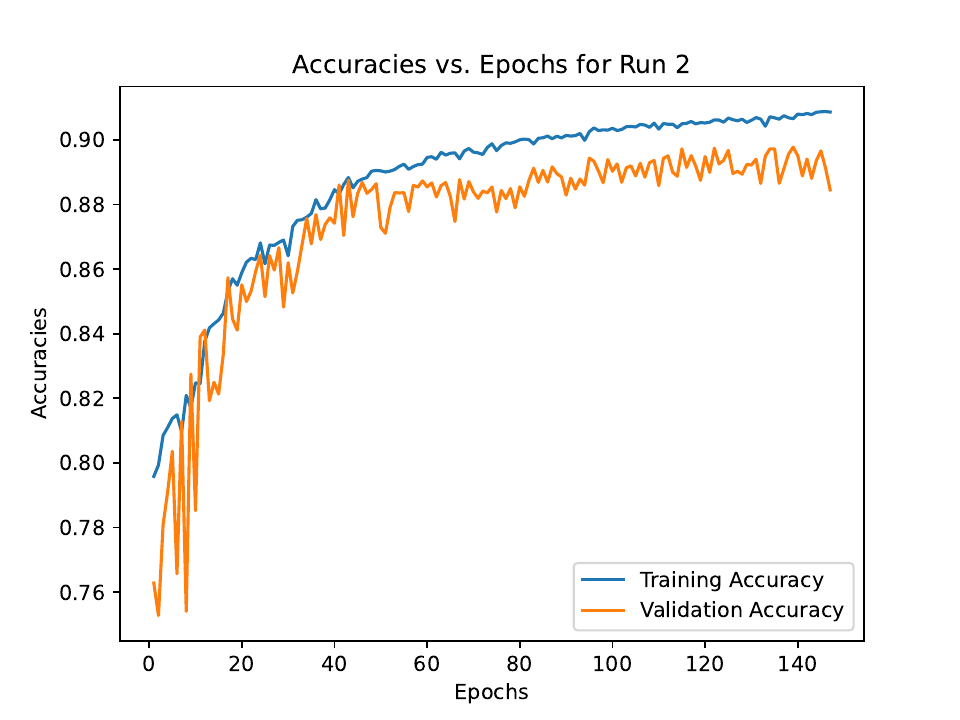}
      }
   \end{minipage}
   \caption{Detailed loss and accuracy curves for our best trial/run (147 epochs) with an overall end train accuracy of 0.909 and test accuracy of 0.896.}
   \label{fig:run_two}
\end{figure*}

During inference, the model first generates a cloud mask for each $256 \times 256$ image of testing data, where the input image goes through the same preprocessing step as described above for training. For each pixel in the image, the model outputs a probability of that pixel having a clear sky. Pixels that have a probability larger than 0.5 are classified as clear-sky and cloudy otherwise. Then, those images are then reconstructed back to full-size masks of dimension $1200 \times 1500$.

\section{Model}

The reference implementation uses a deep U-Net \cite{Ronneberger2015UNetCN} for image segmentation. The architecture of the U-Net allows the model to output a label for each pixel in the input image instead of one label for the entire image itself. To do this, the model first creates a contraction path to get deeper context and then a symmetric expanding path to increase resolution that promotes localization in outputting a prediction.

\section{Experiments}
\label{sec:exp}

\subsection{Resources \& Hardware}
\label{sec:hw}

Our experiments were conducted on New York University's High Performance Computer (HPC) {\em Greene}. We describe the computing resources in more detail and summarize some of their characteristics in Table \ref{tab:hwoverview}. NYU Greene is a general-purpose HPC cluster at New York University that supports a variety of job types and sizes. This includes jobs requiring multiple CPU cores, multiple GPU cards, or terabytes of memory. For this work, we use the V100 series of GPUs from NVIDIA. More details on the specifications of the cluster can be found on the NYU HPC Greene Homepage \cite{www-greene}.

\subsection{Code Modifications}
\label{sec:code}

Work is primarily based upon the open-source code \cite{www-mlcommons-science-github} repository of the MLCommons' cloud-masking benchmark but with additional modifications and functionalities.

\begin{itemize}    
    \item Calculate and display accuracy in the logs.
    \item Fixing issues with improper accuracy calculation in original code.
    \item An early stopping \cite{Caruana2000OverfittingIN} functionality to avoid overfitting.
    \item Checkpoint the weights of the best-performing model.
    \item Control for pseudo-random runs via seed-setting. This allowed us to run more than one experiment and report the variance in the performance.
    \item Functionality that creates temporary YAML files and Slurm job files on the fly so that only one file is sufficient to run all the given jobs in parallel without having to run and specify these jobs separately.
    \item Functionality to save different models in different directories while the jobs are running in parallel. This mitigates the risk of overwriting model objects when the same hyperparameters are used.

\end{itemize}

On NYU's HPC Greene, we train our U-Net model for a total of 200 epochs using the Adam optimizer with default hyperparameter settings and early stopping \cite{Caruana2000OverfittingIN} parameterized with a patience value of 25 epochs---the value was selected from visual inspection and fine-tuning. The potential difficulty of training this model, as observed by unfavorable validation loss curves, is suspected to be due to the usage of Bayesian Masks as the ground truth, which has been observed in the MLCommons Science benchmark \cite{Thiyagalingam2022AIBF} as well. We use a learning rate of $10^{-3}$ with a batch size of $32$. The remaining hyperparameters, such as the crop size, patch size, train test split, etc., are kept the same as those of the reference implementation. We run our experiments over five trials; results are reported in Table \ref{tab: Results}. 
Figure \ref{fig:run_two} shows details for Loss and Accuracy in run two above. The average accuracy achieved over five runs is 0.889.

\subsection{Results}

Figure \ref{fig:all_runs_tr} shows the training loss and accuracy for the five runs; we see that the run with 147 epochs performs the best with the smallest loss and highest accuracy values. Due to early stopping, we get the best model weights at epoch 147, and an accuracy of 0.896. Figure \ref{fig:all_runs_val} shows the same for the validation set while Figure \ref{fig:run_two} shows details for the loss and accuracy trajectories for our best run. The average accuracy achieved over all five runs is 0.889.

\section{Conclusion}

This paper describes improvements to the code and evaluation at NYU Greene for MLCommons Science cloud-masking benchmark. In this work, we contributed to community effort by fixing errors with accuracy and improving logging, by introducing early stopping, checkpoints, and use of seeds, by adding functionality enabling simpler running of parallel jobs. We conducted a benchmark evaluation at NYU HPC Greene cluster and reported on the best performance. We have been keeping the MLCommons team informed about our progress and offered our code for submission by raising a pull request on the MLCommons Science benchmark's GitHub repository. Our code is available at \cite{www-aifsr-github-benchmarking}.

\section*{Contributions}

Varshitha wrote the majority of the code and performed most of the computations at NYU HPC. Laiba managed the development of the project. Varshitha and Laiba wrote the main part of the manuscript. Sergey, Dan, and Banani contributed to project development and manuscript preparation. Sergey oversees and advises the AI for Scientific Research (AIfSR) group. 

We are thankful to Gregor von Laszewski (University of Virginia and MLCommons) and Geoffrey C. Fox (University of Virginia and MLCommons) for their help in the ideation and valuable feedback during the project. 

\printbibliography 

\end{document}